\documentstyle[preprint,aps]{revtex}
\begin{document}
\tightenlines
\draft
\title{On  $\eta$-$\pi$ mixing close to the $\eta$-Helium threshold}

\author{A.M. Green\thanks{e-mail "anthony.green@helsinki.fi"}}
\address{Department of Physical Sciences and Helsinki Institute of Physics\\
P.O. Box 64, FIN--00014 University of Helsinki,Finland}
\author{S. Wycech\thanks{e-mail "wycech@fuw.edu.pl"}} 
\address{Soltan Institute for Nuclear Studies,
Warsaw, Poland} 
\date{\today}
\maketitle
\begin{abstract}           
A $K$-matrix formalism is used to relate the  amplitudes for   
the three reactions 
$pd\rightarrow $ $ ^3$He$\eta$, 
$ \pi^-$ $^3$H$  \rightarrow $ $^3$He$\eta$ and   
$pd\rightarrow$ $^3$H$ \pi^+$.
Free parameters  are fitted to the available 
experimental data and an extrapolation below the 
$ \eta ^3$He threshold is made to see the origin of the  
$\eta ^3$He threshold enhancement. The existence of a virtual --- and not a 
quasi-bound  -- state  finds 
support in the data. The  $K$-matrix permits a discussion of   
 $\eta$-$\pi$ mixing. A  mixing parameter of $ 0.010(5) $, i.e. a  mixing angle 
$\theta =0.6(3)^{o}$,  is extracted from a  best fit to 
 the very recent  
$pd\rightarrow $ $^3$He$\pi^0$ reaction data. 
\end{abstract}                                         

\newpage

\section{Introduction} 
\label{intro}

In this paper we concentrate on the few-body interactions of
$\eta$ mesons in three nucleon systems. These  complement our knowledge on the
$\eta$-nucleon interaction and properties of the $\eta$ meson itself. 
Considerable experimental, phenomenological and theoretical work has been 
devoted to understand  the $ \eta ^3$He system. Early SATURNE experiments
found a large  cross section  for the  $pd\rightarrow  \eta  ^3$He reaction
close to  the $\eta$ threshold,   \cite{MAY96}. 
One interesting feature of this process is an enhancement of the 
meson formation 
amplitude in the few MeV energy region  close to the threshold. 
A similar effect was also noted in the studies of the $pn \rightarrow d \eta$ 
reactions made at CELSIUS,   \cite{CAL98}. Both of these reactions 
indicate the possibility  for  virtual or quasi-bound states to be  
formed in these systems. It is expected that in the deuteron the state of interest 
is a virtual state, whereas  in helium it may also  be virtual but there
  a bound state is not ruled out  , \cite{WIL93}, \cite{WYC95},\cite{BEL95}.
While the  slopes of the formation amplitudes indicate the  existence of
such  states,  more detailed properties   may be found  only with an  extrapolation 
of these  amplitudes below the  $ \eta ^3$He threshold. 
This is made possible with the pion production  experiments 
$pd\rightarrow $ $^3$H$ \pi^+ $  and $pd\rightarrow ^3$He$ \pi^0 $
undertaken by COSY,  \cite{MAG00},\cite{MAG03}.  
These results obtained at backward  angles  supplement the
older SATURNE cross sections 
measured  some distance away from the threshold,  \cite{BER85}. 

In this work we present a multiple-channel $K$-matrix analysis of   $ \eta ^3$He formation. 
In addition to the reactions listed above,  we also include the data  from the 
$ \pi^+$ $^3$H$  \rightarrow $ $^3$He $\eta$ process studied at  Brookhaven,  \cite{PEN89}.  
Altogether the data comprise  17 measurements. The  number of 
important  $K$-matrix parameters, obtained with some minor theoretical input, amounts to 4.  
Unfortunately the system is not strongly constrained, and so  new data are welcome. 

A parallel study is devoted to the effect of   $\eta$-$\pi$ mixing in these systems. 
It has been suggested  in Ref.\cite{MAG00} 
that such a mixing is enhanced by the existence of a $\eta^3$He   bound state. 
This   enhancement  has been found in a subsequent experiment, \cite{MAG03}. 
These two related questions are discussed in this paper, 
which is organised  as follows:
\begin{itemize}
\item  Section \ref{sect2} defines  the mixed states of our system.
\item  Section \ref{sect3} discusses the $K$-matrix method  applied to  the
four-channel 
two body subsystems of the meson-3N  system. A full description, in the isospin symmetric limit, 
within  a zero range approximation requires at least five real $K$-matrix elements.    
Couplings to the open three-body and four-body channels induce  phases.  
Since the present data is not 
sufficient for a full determination of all the relevant  $K$-matrix elements,
a model of $S$(1535) dominance is used to remove some ambiguities.
\item  In section \ref{sect4} the comparison with the experimental data 
indicates  an  attractive  $ \eta ^3$He scattering length and the  existence
of  a $ \eta ^3$He virtual state. Also the question of  $\eta$-$\pi$ mixing  
 close to the  $ \eta ^3$He threshold  is discussed.
\item  In section \ref{Con} some conclusions are made.
\end{itemize}
\section{Mixed states } 
\label{sect2}

Let the isospin invariant states of  $\eta$ and  $\pi^0$ mesons be described by 
$\mid  \bar{\eta} )$  and $\mid  \bar{\pi} )$. 
Due to  some  isospin mixing interaction  $H_m$ these states mix into the 
physical states  
$\mid   \eta  )$  and $\mid   \pi  )$. The relation of these two equivalent
sets of states is 

\begin{equation}
\label{m1}
  \mid   \eta  ) = N [ \mid  \bar{ \eta } )  -   \theta  \mid  \bar{ \pi } ) ]
\ \ \ \ {\rm and} \ \ \ \ 
  \mid   \pi  ) =  N [ \mid  \bar{ \pi } )  +  \theta  \mid  \bar{ \eta } ) ] , 
\end{equation}
where  $\theta$  is a mixing parameter  and  $ N= 1/ \sqrt{1 + \theta^2} $ is the normalisation 
factor. These two sets of states form a complete orthonormal basis with the relations 
\begin{equation}
\label{m3}
 (  \eta  \mid  \pi )  = 0 =   ( \bar{ \eta }   \mid  \bar{ \pi } )
\ \ \ \ {\rm and} \ \ \ \ 
 ( \bar{ \eta } \mid  \pi )  = N \theta  = -  ( \bar{ \pi }   \mid  \eta ) .
\end{equation}
The mixing parameter follows from interactions at the quark level. The latter are due to differences 
in the light quark masses and to electromagnetic effects. For calculations of the mixing angle we refer 
to the review  in Ref.~\cite{mixtheory}.  This angle 
 is usually expressed as  
\begin{equation}
\label{m3c}
  \theta  =  \frac {( \bar{ \eta }  \mid  H_m \mid  \bar{ \pi }) }
{ E_\pi^2 -  E_\eta ^2   }, 
\end{equation}
where $  E_\pi , E_\eta  $  are the meson energies.

The transitions in the few-body systems may be analysed in terms of the scattering matrices $T$  
that lead to the physical $\eta$ and $\pi$ in the final states 
\begin{equation}
\label{m4}
 T( \eta )=  ( p d  \mid  T   \mid \eta ^3He  ) 
\ \ \ \ {\rm and} \ \ \ \ 
 T( \pi )=  ( p d  \mid  T    \mid \pi ^3He  ).
\end{equation}
On the other hand, when discussing the formation processes and final state $\eta$  interactions,
that are supposed to be isospin conserving,  it 
is more convenient to use the isospin basis 
\begin{equation}
\label{m5}
 T( \bar{\eta} )=  ( p d  \mid  T    \mid \bar{\eta} ^3He   ) 
\ \ \ \ {\rm and} \ \ \ \ 
 T( \bar{\pi})=  ( p d  \mid  T    \mid \bar{\pi} ^3He  ).
\end{equation}
The simple relationship  between these amplitudes is 
\begin{equation}
\label{m6}
 T( \eta )=  T(\bar{\eta}) (  \bar{\eta}  \mid  \eta ) + T( \bar{\pi}) 
(\bar{ \pi }\mid   \eta   )\approx T(\bar{\eta})-\theta T( \bar{\pi}) 
\end{equation}
\begin{equation}
\label{m6a}
 T( \pi )=  T( \bar{\eta} )   ( \bar{\eta} \mid  \pi )     +    T( \bar{\pi})
( \bar{ \pi } \mid  \pi )\approx  T( \bar{\pi})+\theta T(\bar{\eta}).
\end{equation}
The physical $T$ matrix elements are used  to describe the experimental data. However, to 
parameterise these  data we are going to use 
the isospin basis amplitudes for the scattering  matrix $T$ and for 
the reaction matrix  $K$.

\section{  $K$-matrix analysis of the $\eta$-$^3$He production process }
\label{sect3}

   In this section, the S-wave scattering matrix $ T$  for the {\it idealised}  
four channel two-body system
$\mid p d )$, $  \mid  ^3$He$\bar{\eta})$, 
 $ \mid $$ ^{3}$He$\bar{\pi}^{0}) $, $\mid  ^3$H$ \pi^+ )$  
is introduced. These channels are denoted by the suffices  $p,\eta,\pi,\pi^+$ 
respectively. The  important couplings to open few-body channels are at 
first forgotten.  The scattering matrix and the basic interactions 
are described and parameterised  in terms 
of a real  $K$-matrix. Next,  the Heitler equation
for the $T$-matrix is solved. This becomes a simple matrix equation 
\begin{equation}
\label{k1}
 T_{j,k}=  K_{j,k} +  i \Sigma_l  K_{j,l}q_{l}T_{l,k},
\end{equation}
where $j,k,l$ are the channel indices 
and $ q_{l} $ is a diagonal matrix of the CM momenta in each channel.    
Here the  region of interest spans 
from about 10 MeV below the $ \eta  ^3$He  threshold to some 10 MeV
above it. For these low energies the  experimental data exists. Also, 
in this region  the $K_{j,k}$ elements are believed to be constant. 
The model used here supposes the 
interactions to conserve  isospin. In this way the $K$-matrix is 
calculated in-between the 
$\mid  \bar{\eta} )$, $\mid  \bar{\pi} )$
and $\mid  \bar{\pi}^+ )$ states. Since the isospin symmetry for the 
$ \pi^+$ is supposedly not violated one has 
$\mid  \bar{\pi}^+ ) = \mid   \pi^+ )$ and, in addition, some simple 
symmetries relate the matrix elements in the 
 $ \mid $$ ^3$He$\bar{\pi}^{0}) $ and $\mid  ^3$H$ \pi^+ )$ 
channels.
In this way the parameters needed to fix the 4$\times$4 $K$-matrix are reduced 
from 10 to  6. 
These are 
$  K_{p,p}, K_{p,\eta},K_{p,\pi},K_{\pi,\eta},K_{\eta,\eta},K_{\pi,\pi}$.
In practice,  only five  of these matrix elements are needed, since the coupling 
to the entrance $pd$ channel is very small and so  one finds $  K_{p,p}$ to be irrelevant. 
Other elements are related by the isospin symmetry: 
$ K_{p,\pi^+}= \sqrt{2}K_{p,\pi};  
\ K_{\pi^+,\eta} =\sqrt{2} K_{\pi,\eta}; \   
K_{\pi^+,\pi^+}=2K_{\pi,\pi}; \ K_{\pi^+,\pi} =\sqrt{2} K_{\pi,\pi}$.

The solution  of the four dimensional Eq.(\ref{k1})  may be brought to a typical 
form 
\begin{equation}
\label{k2}
 T(p,\bar{\eta})=
\frac {A_{p,\eta}} { 1- iq_{\eta} A_{\eta,\eta}}, 
\end{equation}
where the scattering length in the  $\eta ^3He $ system $A_{\eta,\eta}$ is given by 
\begin{equation}
\label{k3}
  A_{\eta,\eta} = K_{\eta,\eta} + 
\frac{i 3 q_{\pi} K_{\pi,\eta}^2 } { 1- i 3 q_{\pi} K_{\pi,\pi}} 
\end{equation}
and the transition length $A_{p,\eta}$ is 
\begin{equation}
\label{k4}
  A_{p,\eta} = K_{p,\eta} +
 \frac{i 3 q_{\pi} K_{p,\pi} K_{\pi,\eta}} { 1- i 3 q_{\pi} K_{\pi,\pi}}.
\end{equation}
For the pion production amplitudes one obtains 
\begin{equation}
\label{k5}
T(p,\bar{\pi})
= \frac{A_{p,\pi}} { 1- iq_{\eta} A_{\eta,\eta}} =
-\frac{1}{\sqrt{2}} T(p,\bar{\pi}^+),
\end{equation}
where
\begin{equation}
\label{k6}
  A_{p,\pi} = \frac{K_{p,\pi}
[1 -i q_{\eta} K_{\eta,\eta}]+ i q_{\eta} K_{p,\eta} K_{\eta,\pi}} 
                                 { 1- i 3 q_{\pi} K_{\pi,\pi}} 
\end{equation}
and the isospin relationship between the  $ \bar{\pi}^+$
and $\bar{\pi}$ is satisfied.
Equations (\ref{k2} - \ref{k6})  contain  expressions which 
change rapidly in   the small $q_{\eta} $ region.
One such term involves  the  scattering length $A_{\eta,\eta}$. 
A  quasi-bound state, if it exists, is given by the condition  
\begin{equation}
\label{k7}
 1- iq_{\eta} A_{\eta,\eta} = 0,
\end{equation}
which is to be satisfied  by a complex momentum  $q_{\eta}^B$. 
 For a large 
$ A_{\eta,\eta}$ this momentum is close to the threshold 
and the factor $ 1 /(1- iq_{\eta} A_{\eta,\eta}) $ induces rapid 
energy dependence of the formation amplitudes in this region. 
This has been  found in the $\eta$ 
production experiment \cite{MAY96}, \cite{WIL93}, but one can 
expect a similar behaviour in other channels. 
However, 
in the pion production amplitudes  another factor arises  
which tends to suppress such an effect. It  is due 
to  the $A_{p,\pi}$ of Eq.(\ref{k6}). There the  existence of a  
quasi-bound state   
is expressed as the dominance of  $K_{\eta,\eta}$.  
Therefore,  if $K_{\eta,\eta}$ were to differ from $A_{\eta,\eta}$
in a significant way, then one would expect a sizable energy dependence 
in the pion formation amplitude. However, as will be seen later, the
indications are that $K_{\eta,\eta}\approx A_{\eta,\eta}$. 
This question  is discussed in the following section.

\subsection{ Estimate of large and small terms in the   $K$-matrix }
\label{sect3a}

The real physical situation involves the coupling of the two body channels  to the 
continuum spectrum of three-body  $NNN, Nd \pi $   and four-body $NNN \pi$ 
channels. The coupling to these systems may be strong and it is not easy 
to calculate. On  phenomenological grounds it requires additional terms 
in the $K$-matrix, so that 
\begin{equation}
\label{k8}
 K_{j,k} \rightarrow   K_{j,k} +   \Sigma_c  K_{j,c} \frac{i  q_{c}} { 1- i  K_{c,c}
q_{c} }  K_{c,k}, 
\end{equation}
where the  summation (an integration) over the continuum  few-body channels $c $ 
is to be performed. 
This equation  induces  complex contributions to the real two channel matrix elements. 
In this subsection we estimate the magnitude of these contributions. 
The phases that arise 
are calculated in terms of  a model or left to a best fit determination.  
The fine tuning 
of the parameters is done in the next section.
The large matrix elements are those related   to the low energy 
$\eta ^3$He  channel. These enter essentially in the form of final 
state interaction factors in the  $\eta ^3$He system. 
One can visualise the relationship between   the $K$-matrix parametrisation 
and a model description 
of the meson formation via  an expression 
\begin{equation}
\label{k9}
T(i,\bar{\eta}) =   \int d {\bf r} d{\bf r'}   \psi_{i}({\bf r'} )
 U_{i,\eta}({\bf r'},{\bf r}) \left[ j_0(q_{\eta} r)   + 
\exp(i q_{\eta} r) \frac{T_{\eta,\eta}}{r} \right].
\end{equation}
Here, $\psi $ denotes a  wave function in the initial channel  $i$ and
$ U_{i,\eta}$ is an operator responsible for  the meson formation. 
The  term in square brackets 
is the final state wave function   expressed in terms of the  
$ \eta $He  scattering matrix $ T_{\eta,\eta} = A_{\eta,\eta}/(1- iq_{\eta}  A_{\eta,\eta}) $.  
Up to terms linear in the final momentum $q_{\eta}$,  one  
obtains  from   Eqs.(\ref{k2}) and (\ref{k9}) 
\begin{equation}
\label{k10}
  A_{i,\eta} = \bar{U}_{i,\eta} \left[1 +  \frac{A_{\eta ,\eta}} { R} \right]. 
\end{equation}
The $ \bar{U}_{i,\eta} $ and $ \bar{U}_{i,\eta}/R $  are  results 
of the integrations in Eq.(\ref{k9}).  The radius  $ R$ reflects  the 
range of  final state interactions and is expected to be 
close to the $^3$He radius. Formula (\ref{k10}) contains effects 
from  all the channels characterised by high intermediate momenta 
i.e.   all   $K$-matrix 
elements other than  $ K_{\eta,\eta} $ or  $ A_{\eta,\eta} $,  
which have already been  specified explicitly.  
Now,  we show that the other  matrix elements are small. Some will be 
needed only to the leading order, whereas others  may be simply neglected. 
To do this we resort  to  the experimental data.

\begin{figure}[ht] 
\includegraphics{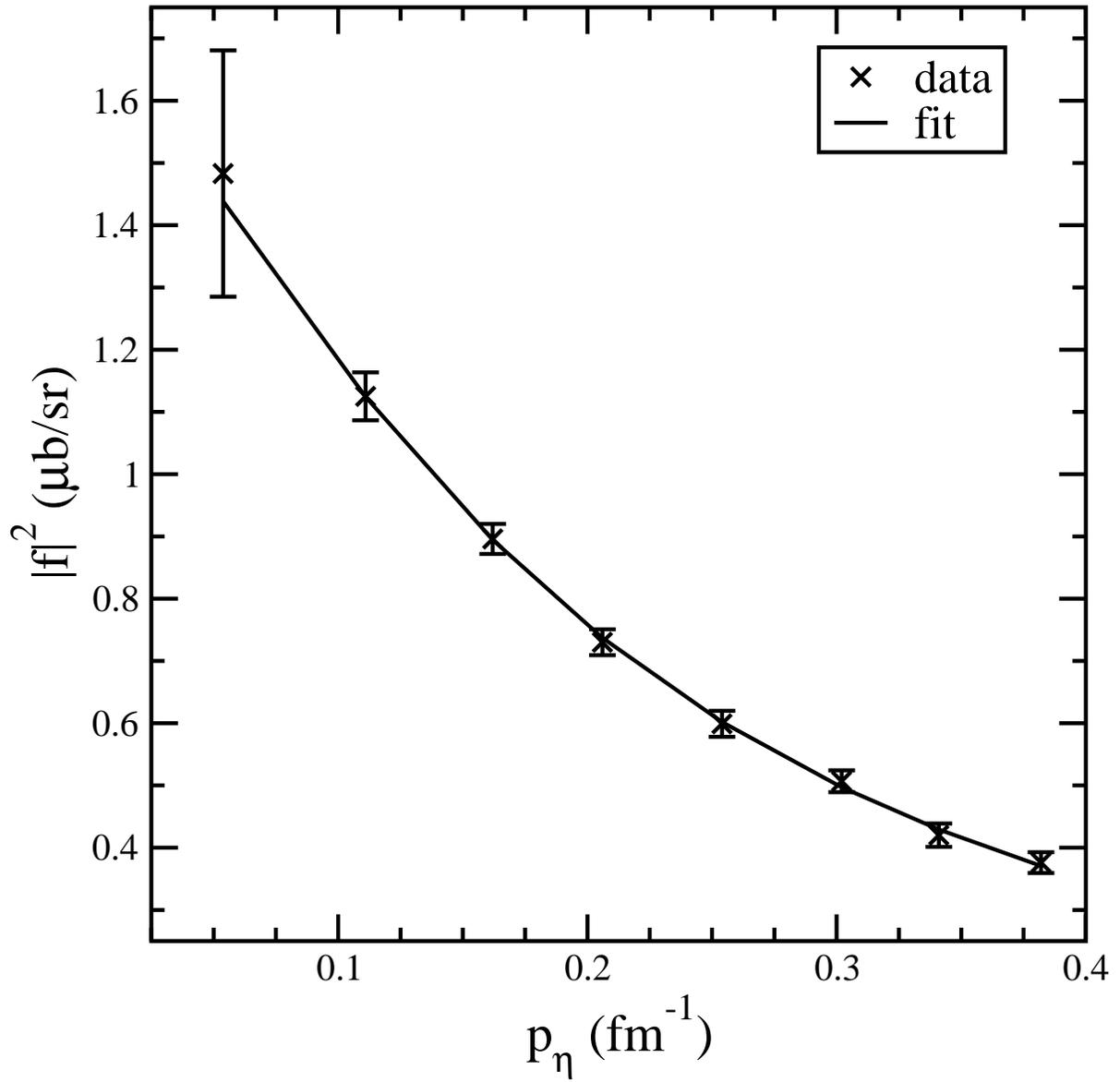} 
\caption{The $pd\rightarrow  \eta  ^3$He data 
(the formation amplitude $\mid f \mid =  \mid T(p,\eta)\mid $  squared ) 
of Ref.~\protect\cite{MAY96}
and the  corresponding  $K$-matrix fit} 
\label{Petafit.} 
\end{figure}

The SATURNE cross section, \cite{MAY96}, for the 
$pd\rightarrow $ $ ^3$He $\eta$ reaction 
is given by   $T(p,\bar{\eta})$   of Eq.~(\ref{k2}). 
These data, given in Fig.~\ref{Petafit.},   fix  
$\mid  A_{p,\eta}\mid $ rather precisely 
to the value of $ 0.013$ fm. On the other hand,  
provided $\mid A_{\eta,\eta}\mid  \approx 5 fm$, 
there is  a whole region of $Re A_{\eta,\eta} $ 
and  $Im  A_{\eta,\eta} $  values that are equally 
likely. For example,the modulus of  Eq.(\ref{k2}) is invariant with respect 
to the sign  change  $Re A_{\eta,\eta} \leftrightarrow -Re A_{\eta,\eta}$. 
These two possibilities 
describe different physics. Large positive values of $Re A_{\eta,\eta}$ 
correspond to a virtual state, an analogue of the NN spin 0 
state at low energies and the singularity of the scattering matrix given 
by Eq.(\ref{k7}) is located in the third quadrant of the 
complex $q_{\eta}$ plane.  The other option, a negative length,  signifies 
a quasi-bound state, which is  analogous to  the deuteron.  One sees that 
the formation cross section  can not distinguish between these two 
possibilities. 

The Brookhaven data, for the  $ \pi^+$ $^3$H$  \rightarrow $ $^3$He $\eta$
process, \cite{PEN89}, permits one to extract two  small 
$K$-matrix elements. These data are described by the amplitude
\begin{equation}
\label{k11}
-\frac{1}{\sqrt{2}} T(\pi,\eta) =  \frac{K_{\pi,\eta} }
{ (1- iq_{\eta} K_{\eta,\eta}) (1- iq_{\pi} K_{\pi,\pi})  - q_{\eta}q_{\pi} K_{\pi,\eta}^2  }, 
\end{equation}
giving    $\mid  K_{\pi,\eta}\mid  \approx 0.07$ fm, a value 
that depends only slightly on the  
choice of  $K_{\eta ,\eta}$.  The coupling of  $^3$He$\eta$ to  
$^3$He($^3$H)$\pi$ 
is rather weak, and an inspection of Eq.(\ref{k3}) tells us  
that its contribution 
to Im$ A_{\eta,\eta} $ is quite small. Therefore, the  Brookhaven 
experiment implies that the $\eta ^3$He 
state decays mainly into three- or four-body systems. Now, both  
$\eta$ formation experiments 
permit a simple description of the two body channels. 
Since it is well established that low energy  $\eta N$
physics is  dominated by the $S(1535)$ resonance we  extend this 
dominance to  few body systems. Thus the channel 
coupling is given by 
\begin{equation}
\label{k12}
  U_{i,j } =   F(q_i) \sqrt{ \gamma_i } \langle 
\frac{1 } { E_{S} - E } \rangle \sqrt{ \gamma_j}   F(q_j)   ,
\end{equation}
where the  $\gamma_i $  couple  the resonance to the 
meson-nucleon channels,  
$\langle \ldots  \rangle$ denotes  a suitable average of 
the resonance propagator over the binding and recoil energies 
and  the $  F(q_j)  $ are the form factors for the 
meson-$^3$He ($^3$H) systems. 
The latter  are expected to be about unity in the $\eta $ channel 
and small in the $\pi$ channels due to the high momenta involved.  
The ratio $ \gamma_{\pi}/\gamma_{\eta}$
may be extracted from  $S(1535)$ decay. Next, Eq.(\ref{k12}) 
--- when combined with the $(\pi,\eta)$ data  --- yield  
$\mid  K_{\pi,\pi}\mid  \approx 0.001$ fm,  a negligible value. 
There is another consequence of Eq.(\ref{k12}), the phase of  
$ U_{\pi,\eta} $ is given by the phase of the $S(1535)$ propagator. 
For low energy $\eta ^3$He  scattering, the relevant energies in 
the meson-nucleon system 
fall  well below the resonance. Therefore, the dominant mode of decay 
is closed and  $ U_{\pi,\eta} $ is almost real. 
The uncertainty in the relative sign in the coupling constants  
$\sqrt{ \gamma_i}$ and $\sqrt{ \gamma_j} $ may be removed 
by the $S(1535)$ state wave function, where the
SU(3) coupling mechanism of Ref. \cite{RIS01} gives  a positive
sign for $  K_{\pi,\eta} $.

We are now ready  to study  the  $pd\rightarrow$ $^3H \pi^+$   reaction,
\cite{MAG03}. 
Given a  small value of $  K_{\pi,\eta} $, alongwith a  negligible 
$ K_{\pi,\pi} $, then
Eqs. (\ref{k5}, \ref{k6}) and the experimental data from ref.~\cite{MAG03} 
yield a crude estimate of   $\mid  A_{p,\pi}\mid   \approx 0.00021$ fm.  

\section{ Fitting the  $K$-matrix to experimental data }
\label{sect4}

The data used  consists of  8 measurements of the  $ ^3$He$\eta $ cross section in the 
threshold region   \cite{MAY96},  4 measurements of the $^3$H$ \pi ^+ $ 
cross section   \cite{MAG03}, 
 4  measurements of the $ ^3$He$ \pi ^0 $ cross section   \cite{MAG03}, and one result  
for  the  $ \pi^+$ $^3$H$  \rightarrow $ $^3$He $\eta$ reaction \cite{PEN89}.
We limit the available data  to the low energy 
$ \eta $ region to stay with the approximation of a constant 
$K$-matrix.    The first two reactions yield absolute values 
of the $K_{p, \eta}$ and $ K_{p,\pi} $  matrix elements. However, 
the relative phase of these  has to be left to an experimental 
determination.  We set 
\begin{equation}
\label{k13}
K_{p, \pi} =  \frac {K_{p ,\eta}} {\omega} \exp(i\psi),
\end{equation}
where 
$ \omega = \mid K_{p,\eta}/K_{p,\pi}\mid =5.54(50) $  is well 
determined from the  $ \eta  $ and $ \pi ^+ $  formation 
experiments.  The phase  $\psi $ 
and the   $\eta$-$\pi$ mixing angle $\theta $ are  free parameters. 
To elucidate the interference pattern in the equations of the 
previous section,  the $\pi$ formation amplitudes  are 
now presented  in a simplified form. Forgetting  an irrelevant  overall 
phase,  up to  terms linear  
in  small  $K_{\pi,\eta}$  one has  
\begin{equation}
\label{k14}
-\frac{1}{\sqrt{2}} T(p,\bar{\pi}^+)= 
K_{p,\pi} [ \exp(i\psi) + 
\frac{iq_{\eta} K_{\pi,\eta}} { 1- iq_{\eta} A_{\eta,\eta}}  \omega ]
\end{equation}
\begin{equation}
\label{k15}
T(p,\bar{\pi}^0)= 
K_{p,\pi} [ \exp(i\psi) + 
\frac{iq_{\eta} K_{\pi,\eta}+\theta} { 1- iq_{\eta} A_{\eta,\eta}} \omega]. 
\end{equation}
In these equations  the two $K$-matrix parameters are real.
A positive  sign for  $K_{\pi,\eta} $ 
is preferred by  the $S(1535)$ dominance and also by the best fit to 
the experimental data.   

Using the {\bf Minuit} minimization package, an overall best fit 
search  to  the data  yields: 
 
\noindent $ K_{p,\eta} = 0.0115(9)$ fm, $K_{p,\pi}= 0.00207(3)$ fm, 
$K_{\pi,\eta}= 0.067(11)$  fm, $K_{\pi,\pi }= 0$ ,

\noindent $K_{\eta ,\eta}=  4.24(29)+i 0.69(81)$ fm, 
$ A_{\eta,\eta}= 4.24(29) +i 0.72(81)$ fm, 

\noindent $\psi= 4.14(27)$
 and 
$ \theta = 0.010(0.005)=  0.6(3)^{o}$. The small difference between 
 $Im \ K_{\eta ,\eta}$ and  $Im \ A_{\eta ,\eta}$ is due to the explicit
inclusion of $K_{\pi,\eta}$. As discussed in the last section, 
the large error in $Im \ A_{\eta ,\eta}$ arises since  the $\eta$ 
formation cross section 
is not restrictive on the values of $ A_{\eta ,\eta}$. 
The real part of the $ ^3$He $ \eta $ scattering length is seen to be 
large and positive --- signalling 
the existence of a virtual-state in this system.

\subsection{  $\eta$-$\pi$ mixing }
\label{sect4a}

\begin{figure}[ht] 
\includegraphics{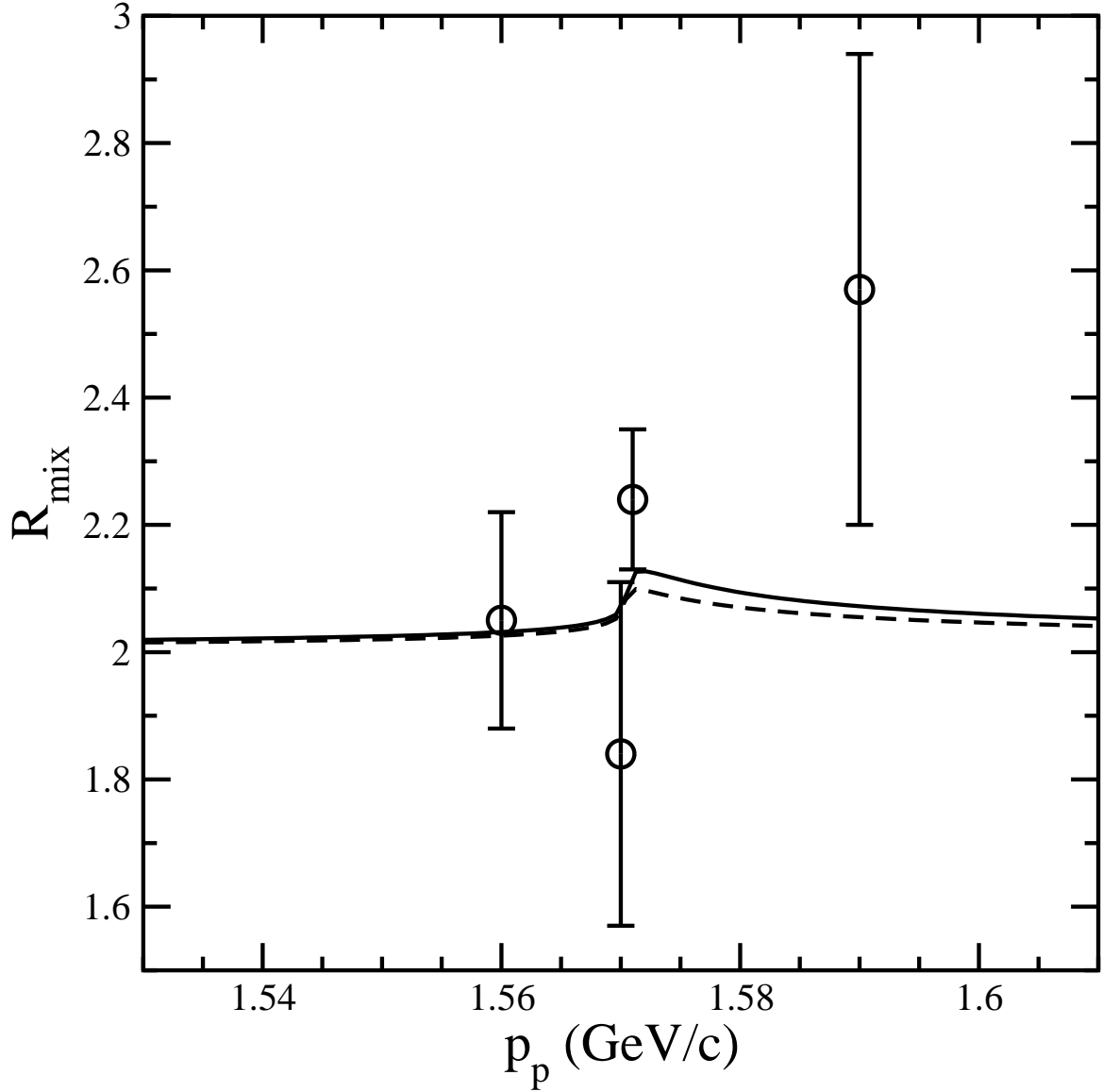} 
\caption{The ratio $R_{mix}= \mid T(\pi ^+) \mid^2 /\mid T(\pi ^0) \mid^2$
calculated with the mixing  angle $\theta = 0.010$.
\newline The experimental results are from  COSY  \protect\cite{MAG03}.
The  dashed   curve is the $K-$matrix fit where the error bars of the 
$pd\rightarrow$ $^3$H$ \pi^+$ and $pd\rightarrow $ $^3$He$\pi^0$  data
have been doubled in an attempt to simulate systematic errors in that data.
These results include the correction factor $\lambda =1.04$ discussed in
Subsect.~\protect\ref{sect5}} 
\label{Ratio} 
\end{figure}

\begin{figure}[ht] 
\includegraphics{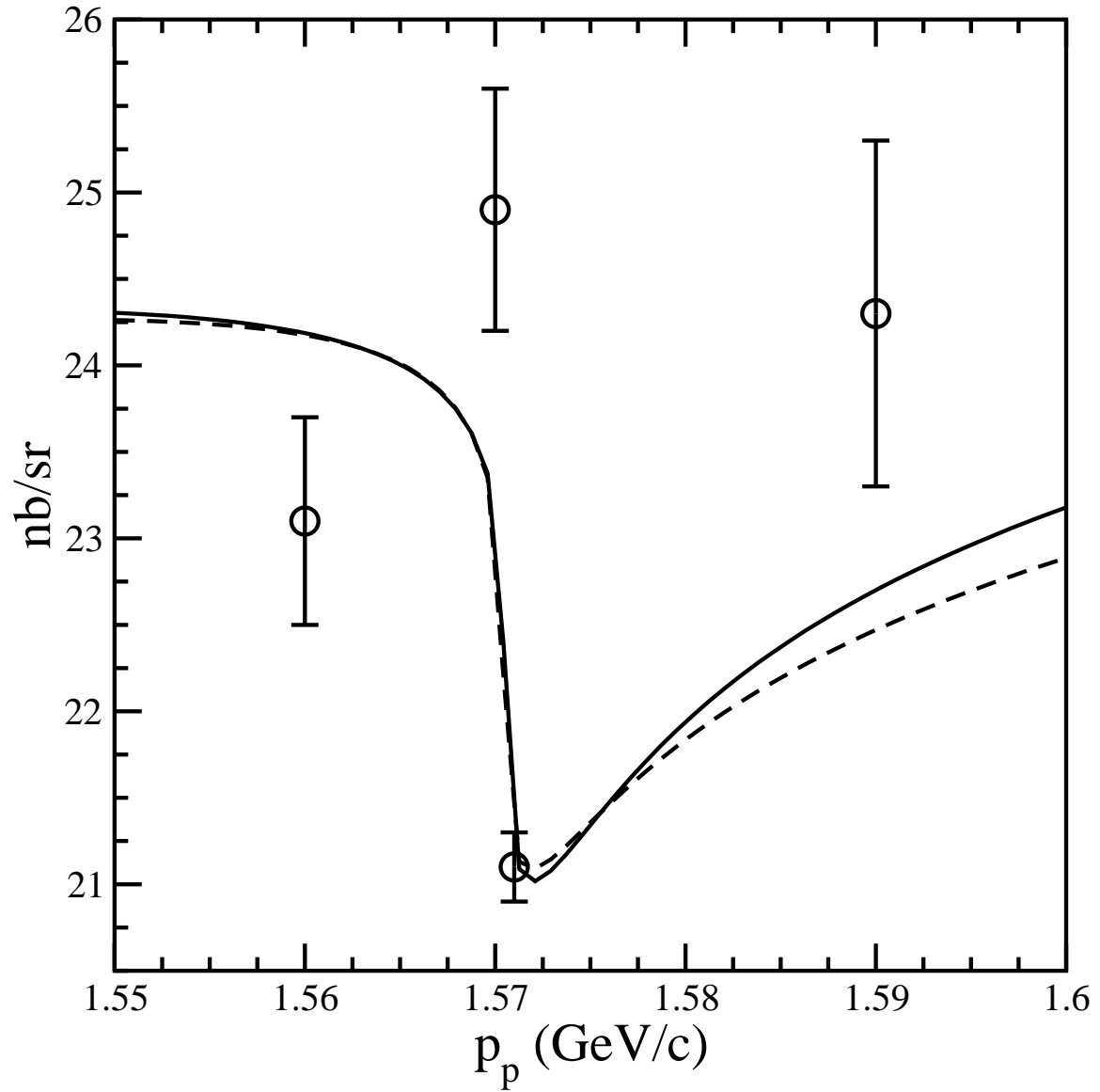} 
\caption{The $pd\rightarrow $ $^3He \pi^0$ cross section calculated with the
mixing  angle $\theta = 0.010$ and $\lambda =1.04$.
\newline The experimental results are from  COSY  \protect\cite{MAG03}. } 
\label{Pi0Xsection} 
\end{figure}

The idea behind the   detection  of $\eta$-$\pi$ mixing  
at COSY  was  to exploit the  ratio of 
the charged and neutral pion cross sections   
$R_{mix}$, \cite{MAG00}. 
According to  isospin invariance,  $R_{mix}$ should equal 2. 
However, the mixing  induces corrections such that   
\begin{equation}
\label{t1}
R_{mix} \equiv \frac { \mid T(\pi^+) \mid^2} { \mid T(\pi ^0) \mid^2} = 
\frac {2  \mid T(\bar{\pi}) \mid^2}  {\mid T(\bar{\eta})(\bar{\eta}\mid \pi) + 
T(\bar{\pi})(\bar{\pi}\mid \pi)\mid^2}  
= \frac {2} {N^2 \mid 1 +  \theta T(\bar{\eta}) / T(\bar{\pi}) \mid^2} .
\end{equation}
There is an additional reason for studying the ratio $R_{mix}$,  rather than the 
separate cross sections, since in this 
way most of the  systematic errors are removed. Because of that we 
limit our discussion to the data obtained 
in one laboratory. 
The  amplitudes obtained in  the previous section  
and  $\theta =  0.010(5) $  [$\theta =  0.6(3)^{o}] $   reproduce 
the  trends in the measured values to a fair  degree 
as is shown in Table~\ref{table1} and  Fig.~\ref{Ratio}.  
In  the latter, as a crude attempt to estimate systematic errors in
the two measured cross sections, the effect of doubling the experimental
error bars is shown as a  dashed line. This is seen to have a minor effect.
The data and overall fit from the model display   a maximum just above
the $\eta$-threshold. However, in the data  a minimum 
below the threshold is also  indicated,  but it is not 
reproduced here. 
For comparison, in Fig.~\ref{Pi0Xsection}, the
$pd\rightarrow $ $^3He \pi^0$ cross section calculated with a mixing  
angle of $\theta = 0.010$ is plotted against  the experimental results 
 \protect\cite{MAG03}.

The last step in this analysis has been  a more detailed  extraction of 
the $\eta$-$\pi$ mixing parameter  $\theta $. 
Notice that  Eqs.(\ref{k14}) and (\ref{k15}), which are essential 
for that procedure,  
 have a very simple structure at the threshold, $q_{\eta} =0 $. 
This point determines the unknown phase $\psi$. 
We find  two basic solutions that differ by the sign of  
$\theta $, where  
the negative sign is ruled out on  physical grounds. 
The actual value of  $\theta $ is then  extracted  by  
the  energy dependence of the ratio $R_{mix}$ and 
the  energy independence  in  $\mid T(p,\bar{\pi}^+) \mid$ 
below the threshold indicated in Table I.
However, the best fit parameters are not well determined and 
their errors are large.
The need for more precise and more numerous data is evident.

The value of the mixing parameter  obtained here 
is smaller than  the  $\theta = 1.5(4)^{o} $  extracted from 
the  $ \pi d \rightarrow \eta NN $
reaction, \cite{TIP01}.  
Theoretical calculations  yield  values in the region of 
$(0.75-0.85^{o} )$  \cite{PIE93} - \cite{LEU96}, although 
angles twice as
large have also been suggested, \cite{BAG90}.

\subsection{ Discussion of corrections }
\label{sect5}

So far we have assumed that all the isospin violation is due to the
$\eta$-$\pi$ mixing. However, there are other sources of  such a 
violation. These can enter in two ways. 
\begin{itemize}
\item The two reactions
$pd\rightarrow$ $^3$H$ \pi^+$ and $pd\rightarrow $ $^3$He$\pi^0$  
need not be exactly in the ratio 2 to 1 as is assumed 
in Eq.~(\ref{t1}),
since they contain explicit isospin violation effects such as the 
difference between the  $^3$He and  $^3$H nuclear wavefunctions, 
Coulomb interactions and different  meson momenta.
This effect is  analysed in terms of an additional  free 
parameter  $\lambda$, by assuming 
$\mid T(\bar{\pi}^+) \mid^2/  \mid T(\bar{\pi}^0) \mid^2 $ 
 to be $2 \lambda$. The best fit 
value of $\lambda = 1.04(5)$ is found to improve  our 
$R_{mix}$ ratio in Fig. 3. In this way, the
mixing parameter is reduced 
 to $\theta = 0.007(5)$. However, this procedure  
uses  a very limited data base. An extension of  
the data could possibly lead  to a different value of $\lambda$. 

On the other hand  qualitative theoretical arguments in Ref.~\cite{Kohler}
do seem to suggest a value of $\lambda $ that is greater than
unity ($\approx 1.10$).
One conclusion from this type of overall renormalisation is that 
the error in $\theta $  
could well be  larger. 
  
\item  An additional isospin 
violation effect in the $pd\rightarrow $ $ ^3$He$\eta$ 
reaction could also arise from $\rho $-$\omega$ mixing. 
However, since our approach is based on  
phenomenological $K$-matrix parameters such an effect  would not change  
directly the present determination of $\theta $.
Presumably this would contribute to an overall normalisation correction
and so is taken into account by the above $\lambda$ correction.
 
\end{itemize}
The conclusion is that, eventhough these two effects could be at 
a 10\% level, they do not lead to any dramatic effect at the $\eta$-
threshold and so are  incorporated in the multiplicative factor $\lambda$.

In addition to the above corrections, it should be remembered that
Eq.~(~\ref{t1}) is written down for S-waves under 
the assumption that 
these dominate in the backward scattering, whereas 
a more correct expression would involve the effect 
of higher partial waves. 
That possibility was incorporated by simply adding an  additional  
"background" contribution as a complex constant ($c$)   to $ T(p,\bar{\pi})$.
However, the best fit 
procedure indicated that $c$ was very small and so ruled out any 
significant contribution of this kind.
\section{Conclusion}
\label{Con}
The $K-$matrix formalism developed here is able to account for the
structure seen at the $\eta$-threshold in the experimental ratio
$\mid T(\bar{\pi}^+) \mid^2/  \mid T(\bar{\pi}^0) \mid^2 $
 as a manifestation of $\eta$-$\pi$ mixing. It also enables
an estimate of 0.010(5) to be made of the mixing angle.
Unfortunately, at present this estimate has a large uncertainty,
which could be significantly reduced  by the removal 
of several uncertain systematic effects in the available
experimental data. This clearly exposes the need for
more precise data over the energy range covering the $\eta$-threshold. 
Such data should be detailed within the range $p_{lab}$= 1.55 -- 1.59 GeV.
In addition, some data points further from the threshold would be very 
valuable in order to study the non-threshold value of 
$\mid T(\bar{\pi}^+) \mid^2/  \mid T(\bar{\pi}^0) \mid^2 $ and so tie
down more precisely model parameters such as $\lambda$.

\vskip 2cm
  
\noindent{\bf Acknowledgments}

Hospitality of the Helsinki University Department of Physical Sciences and
the Helsinki Institute of Physics is acknowledged by S.W. There, 
most part of this work was carried out.
We wish to thank Andrzej Magiera 
for his collaboration and advice and Dan Olof Riska and Christoph Hanhard for
discussions.  
This project is financed by the  
Academy of Finland contract 43982, the  KBN grant 5P03B04521
and 
the European Community Human Potential Program HPRN-CT-2002-00311 EURIDICE.

\begin{table}
\caption{  The experimental and calculated amplitudes    
$\mid T(\pi ^+) \mid^2 [10^{-7} fm^{2}] $ for the  $\pi ^+$ production. 
Other columns  give ratios   $R= \mid T(\pi ^+) \mid^2 /\mid T(\pi ^0) \mid^2$   
calculated with a mixing angle of $\theta = 0.010$.
The experimental results are from  COSY  \protect\cite{MAG03}. 
The first column gives the proton laboratory momentum in GeV.}
\vskip 0.2cm
\begin{tabular}{lcccc}
$p_{lab}$  & $\mid T(\pi ^+) \mid^2_{exp}$ & $\mid T(\pi ^+) \mid^2_{calc}
$ & $ R_{exp}$  & $ R_{calc}$ \\ \hline

 1.560          &  47.4(3.9)   & 49.7     & 2.05(0.17)         &  2.04     \\
 1.570          &  45.8(6.6)   & 48.8     & 1.84(0.27)         &  2.08      \\
 1.571          &  47.5(2.2)   & 46.0     & 2.24(0.11)         &  2.17      \\
 1.590          &  62.6(8.6)   & 47.6     & 2.57(0.27)         &  2.11      \\
 \end{tabular}
\label{table1}
\end{table}

\vskip 3.0cm

{\bf Figure Captions}
\vskip 0.5cm

Fig.1 The $pd\rightarrow  \eta  ^3$He data 
(the formation amplitude $\mid f \mid =  \mid T(p,\eta)\mid $  squared ) 
of Ref.~\protect\cite{MAY96}
and the corresponding $K$-matrix fit.

\vskip 0.5cm

Fig. 2 The ratio $R_{mix}= \mid T(\pi ^+) \mid^2 /\mid T(\pi ^0) \mid^2$
calculated with the mixing  angle $\theta = 0.010$.
\newline The experimental results are from  COSY  \protect\cite{MAG03}.
The dashed curve is the $K-$matrix fit, where the error bars of the 
$pd\rightarrow$ $^3$H$ \pi^+$ and $pd\rightarrow $ $^3$He$\pi^0$  data
have been doubled in an attempt to simulate systematic errors in that data.
These results include the correction factor $\lambda =1.04$ discussed in
Subsect.~\ref{sect5}.

\vskip 0.5cm

Fig. 3 The $pd\rightarrow $ $^3He \pi^0$ cross section calculated with the
mixing  angle $\theta = 0.010$ and $\lambda =1.04$. 
The experimental results are from  
COSY  \protect\cite{MAG03}.

\newpage

\newpage 

\end{document}